\documentclass[aps,twocolumn,showpacs]{revtex4}
\usepackage{graphicx}
\usepackage{dcolumn}
\usepackage{bm}
\begin{document}

\title{Multiple bound states in scissor-shaped waveguides}
\author{Evgeny N. Bulgakov$^{a}$, Pavel Exner$^{b,c}$,
 Konstantin N.Pichugin,$^{a}$ \\ and Almas F.Sadreev$^{a,d}$ \\
 $^a$ Institute of Physics, 660036 Krasnoyarsk, Russia \\
 $^b$ Nuclear Physics Institute, Czech Academy of Sciences, 25068
 \v{R}e\v{z} \\
 $^c$ Doppler Institute, Czech Technical University,
 B\v{r}ehov\'{a} 7, \\ 11519 Prague, Czechia \\
 $^d$ Department of Physics and Measurement Technology, \\
 Link\"{o}ping University, 581 83 Link\"{o}ping, Sweden}
\begin{abstract}
\noindent We study bound states of the two-dimensional Helmholtz
equations with Dirichlet boundary conditions in an open geometry
given by two straight leads
of the same width which cross at an
angle $\theta$. Such a four-terminal junction with a tunable $\theta$
can realized experimentally if a right-angle structure is filled
by a ferrite. It is known that for $\theta=90^o$ there is one proper
bound state and one eigenvalue embedded in the continuum. We show
that the number of eigenvalues becomes larger with increasing
asymmetry and the bound-state energies are increasing as functions
of $\theta$ in the interval $(0,90^o)$. Moreover, states which are
sufficiently strongly bent exist in pairs with a small energy
difference and opposite parities. Finally, we discuss how with
increasing $\theta$ the bound states transform into the quasi-bound
states with a complex wave vector.
\end{abstract}

\pacs{42.25.Bs,03.65.Ge, 73.40.Lq,03.75.Be}
\maketitle

\section{Introduction}

The question of a possible existence of modes trapped in open
two-dimensional systems has been a classic in the theory of
waveguides; trapped modes due to particular boundary conditions
were studied already half a century ago \cite{ursell}. However,
only much later it was realized that the introduction of bends and
crossings into waveguides gives rise generally to confined states,
or bound states which exist below the cutoff frequency for the
waveguide \cite{roukes, timp, exner1, exner2, schult, peeters,
exner3, exner4, avishai, goldstone, carini, carini1, carini2,
bulgakov, bulgakov1}. The existence of such states has both
theoretical significance and implications for possible
applications. They have been subsequently discussed in many
papers, in addition to those mentioned above we refer to the
monograph \cite{londergan} and the bibliography there.

In this paper we consider a system of two straight waveguides of
the same width $d$ which cross at a nonzero angle $\theta$. The
right-angle case was of one of the first examples where the
binding was studied. Schult, Ravenhall, and Wyld \cite{schult}
showed the existence of two bound states. One of them is a true
bound state at the energy $0.66 (\pi/d)^2$ in the natural units,
while the other one at $3.72 (\pi/d)^2$ is embedded into the
continuum and does not decay due to the symmetry. The latter
corresponds to the single bound state in an L-shaped tube of width
$d/2\,$ \cite{exner2}. Our aim is to show how the spectrum of such
a junction, which we will call for the sake of brevity
``scissors'' in the following, changes as the angle $\theta$
varies over the interval $(0,90^o)$.

We will show that as we go further from the cross symmetry if the
right-angle structure, new bound states emerge from the continuum.
In strongly skewed junction corresponding to a small $\theta$ there
are many of them. The mechanism responsible for their existence is
the same as for the bound states in sharply broken tubes studied
theoretically and experimentally in \cite{avishai, carini}, namely
a long part of the junction where there transverse contribution to
the energy is substantially lower than $(\pi/d)^2$. In the present
case, however, the system has a mirror symmetry with respect to
the axis of the complement angle $180^o-\theta$ and the bound states
exist in pairs corresponding to different parity. We will show
that as the angle $\theta$ diminishes and the states become strongly
bound, the energy gap between the even and odd member of the pair
vanishes exponentially fast. We also study the behavior around the
critical values $\theta_c$ where the bound states emerge from the
continuum. Our numerical analysis shows that the binding energy of
the weakly-coupled states behaves as
correction
$\approx \pi^2-\gamma
(\theta_c-\theta)^2$

\section{Ferrite filled microwave waveguides as a way to vary the
angle of the scissors}

Before analyzing the scissor spectrum, let us discuss how such a
structure can be realized experimentally as a microwave device. It
is no problem, of course, to build crossed waveguides in the ways
explained in \cite{londergan}. However, in such a setting it is
not easy to vary the geometry continuously. Our point here that
this goal can be achieved with a structure of a fixed angle if the
latter is filled by a ferrite with an axial magnetic anisotropy
and an external magnetic field is applied. We will show that this
leads to an effective angle controlled by the field strength,
following the idea which was first applied to the equivalence
between a ferrite-filled squared resonator with an external
magnetic field and a field-free rhombic polygon \cite{bulgakov2}.

To explain the mechanism of this equivalence we begin with the
Maxwell equations which in the presence of a material have the
form
\begin{eqnarray}
\label{maxwell} && \nabla\cdot{\bf E} = \nabla\cdot{\bf B}
=0,\nonumber \\ && \nabla\times{\bf E} = -ik{\bf B},\quad
\nabla\times{\bf H} = ik{\bf E},\nonumber\\  && {\bf
B}=\hat{\mu}{\bf H},
\end{eqnarray}
where ${\bf E}$ is the electric field, ${\bf H}$ is the magnetic
field, ${\bf B}$ is the magnetic induction, $k=\omega/c$ and
$\omega$ is an eigenfrequency with the wave vector $k$. We suppose
that the material has a magnetic anisotropy corresponding to an
anisotropic permeability $\hat\mu=1+4\pi\hat\chi$ with
\cite{lax}
\begin{equation}
\label{chi} \hat\chi=\left(\matrix{\chi_{xx} &  \chi_{xy} &  0\cr
                           \chi_{xy} & \chi_{yy} & 0\cr
                           0 &  0  & 0\cr}\right),
\end{equation}
where
$$ \chi_{xx}=\frac{g\Omega_1 M_0}{\Omega_1\Omega_2
-\omega^2},\quad \chi_{yy}=\frac{g\Omega_2 M_0}
{\Omega_1\Omega_2-\omega^2} $$
and
Eq(3)
\begin{equation} \label{mixchi}
 \chi_{xy}= -\chi_{yx}=\frac{i\omega g\Omega_1
M_0}{\Omega_1\Omega_2-\omega^2}.
\end{equation}
Here $g$ is the magnetomechanical factor, $M_0$ is the
magnetization of the material,
\begin{eqnarray}
\label{omega12} \Omega_1=gM_0 \left( \frac{{\bf M}_0 {\bf
H}_0^{(i)}}{M_0^2} + \frac{gK_a}{M_0}\cos^2\Psi
\right),\nonumber\\ \Omega_2=gM_0 \left( \frac{{\bf M}_0 {\bf
H}_0^{(i)}}{M_0^2} + \frac{gK_a}{M_0}\cos2\Psi \right)
\end{eqnarray}
and $K_a$ characterizes the anisotropy type: it is an easy plane
anisotropy for $K_a>0$ and an easy axis anisotropy for $K_a<0$. In
what follows we suppose that the material has an easy plane
magnetic anisotropy, $K_a>0$, in which case the intrinsic magnetic
field is equal to
$$ {\bf H}={\bf H}_0-4\pi M_{0z}. $$
On the relations (\ref{omega12}) $\:\Psi$ is the angle between the
anisotropy axis ${\bf N}$ and the magnetization ${\bf M}_0$. We
choose the latter to coincide with the $z$ axis along the
magnetization, while the $x$ axis lies in the plane spanned by the
vectors ${\bf N}$ and ${\bf M}_0$.

\begin{figure}[t]
\includegraphics[width=.4\textwidth]{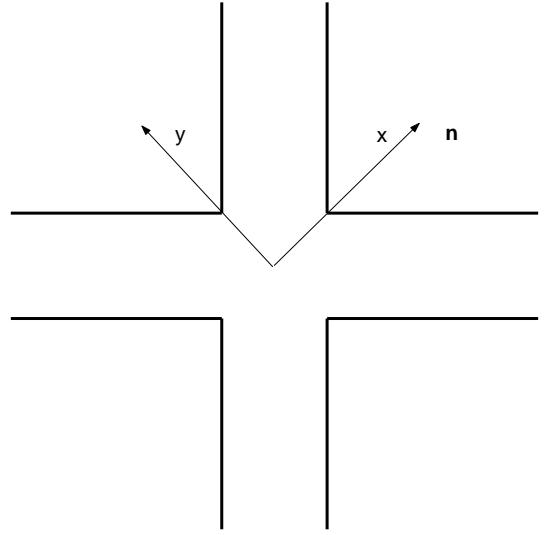}
\caption{Schematical view of the cross-bar resonator (scissors
with $\theta = 90^o$)  filled with ferrite where ${\bf M}$ is the
magnetization of ferrite and ${\bf N}$ is the anisotropy field.}
\label{scisfig1}
\end{figure}

In the simplest case of an easy plane magnetic material we have
${\bf M}_0 \perp {\bf N}$ and ${\bf M}_0 \parallel {\bf
H}_0^{(i)}$ with $\Psi=\pi/2$, so we obtain from (\ref{omega12})
\begin{eqnarray}
&& \label{omega120} \Omega_1=g(H_0 -4\pi M_0),\nonumber\\ &&
\Omega_2=g(H_0-4\pi M_0)+gK_a M_0.
\end{eqnarray}
This is shown in Fig.~\ref{scisfig1} where the $z$ axis is
perpendicular to the plane of the waveguide.
We seek a
two-dimensional solution of the Maxwell equations (\ref{maxwell})
shown at this figure in the form ${\bf E}(x,y)=E(x,y){\bf e}_z$.
The fields ${\bf B}, {\bf H}$ lay in the plane $x, y$ and depend
on $x, y$ too. Then the first equation is satisfied, while the
third one gives
  \begin{eqnarray}
\label{maxwell3}
   -ikB_x=\frac{\partial E_z}{\partial y}, \quad
-ikB_y=-\frac{\partial E_z}{\partial x},
  \end{eqnarray}
and finally, the fourth Maxwell equation can be rewritten as
\begin{equation}
\label{maxwell4} ikE_z=\frac{\partial H_y}{\partial
x}-\frac{\partial H_x}{\partial y}.
\end{equation}
Using the explicit form of the permeability given by (\ref{chi})
we get
 $$
 \mathbf{B}=\left(\matrix{\mu_{xx} & \mu_{xy} & 0 \cr
  \mu_{yx} & \mu_{yy} & 0 \cr
   0 & 0 & 1\cr}\right)\left(\matrix{H_x \cr
   H_y \cr
    H_z\cr} \right).
$$
Combining this with Eq.~(\ref{maxwell3}) we obtain
\begin{equation}
\label{HE} \left(\matrix{H_x\cr H_y\cr}\right)=\frac{1}{D}
\left(\matrix{\mu_{yy} & \mu_{xy}\cr \mu_{yx} & \mu_{xx}}\right)
\left(\matrix{\frac{i}{k}\frac{\partial E_z}{\partial y}\cr
\frac{-i}{k}\frac{\partial E_z}{\partial x}\cr}\right),
\end{equation}
where we have denoted
 $$ D=\mu_{xx}\mu_{yy}-\mu_{xy}\mu_{yx}. $$
Substituting Eq.~(\ref{HE}) into Eq.~(\ref{maxwell4}) we obtain
\begin{equation}
\label{mix} \mu_{xx}\frac{\partial^2{E_z}}{\partial{x^2}}+
\mu_{yy}\frac{\partial^2{E_z}}{\partial{y^2}}+
(\mu_{xy}+\mu_{yx})\frac{\partial^2{E_z}}{\partial{x}\partial{x}}+
D k^2 E_z=0.
\end{equation}
The key observation is that the mixed derivatives in the last
equation can be eliminated by the coordinate transformation
\begin{equation}
\label{trans} \left(\matrix{x' \cr y' \cr }\right)=
\left(\matrix{-\frac{\sqrt{\mu_{xx}\mu_{yy}-(\mu_{xy}+\mu_{yx})^2/4}}{\\mu_{xx}}
&0 \cr
   -\frac{\mu_{xy}+\mu_{yx}}{2\mu_{xx}} & 1 \cr} \right)
\left(\matrix{x \cr y \cr }\right)
\end{equation}
which allows us to cast Eq.~(\ref{mix}) into the following simple
form
\begin{equation}
\label{helmholz} \nabla^2 E_z+\lambda k^2 E_z=0,
\end{equation}
where
\begin{equation}
\label{lambda0} \lambda=\sqrt{\frac{\mu_{yy}}{\mu_{xx}}}\,;
\end{equation}
we have taken into account that $\mu_{xy}+\mu_{yx}=0$ holds in
accordance with (\ref{mixchi}).

The transformation (\ref{trans}) defines a relation between a
right-angle cross structure and a skewed one with an angle defined
by $\lambda$. It is too daring, however, to speak about a full
equivalence, because it is clear from the formulas expressing the
elements of (\ref{chi}) that the angle depends on the
eigenfrequencies involved. Let us ask under which conditions this
dependence of the geometrical factor (\ref{lambda0}) can be
suppressed.
Substituting into (\ref{lambda0}) the expressions
(\ref{omega120}) we get
\begin{equation}
\label{lambda2} \lambda^2=\frac{\Omega_1\Omega_2 -\omega^2+4\pi
g\Omega_2 M_0} {\Omega_1\Omega_2 -\omega^2+4\pi g\Omega_1 M_0}.
\end{equation}
Following to \cite{bulgakov2} we can simplify this expression if
to assume that
\begin{equation}
\label{KgMH} gK_a/M_0 \gg \max\{ gH_0,  4\pi gM_0, \omega \}.
\end{equation}
For typical ferrites $K_a \sim 10^6\: {\rm erg/cm}^3$ and $4\pi
M_0 \sim 100\: {\rm Oe}\:$ \cite{lax}.
Taking the
magnetomechanical factor $g\sim 10^7 sec^{-1}Oe^{-1}$ we obtain
from(\ref{KgMH}) $H_0 \ll 10^4\: {\rm Oe},\; \omega \ll 10^{11}$
what would require very wide waveguides of a width $d\sim 10\:
{\rm cm}$. However, there are ferrites with $K_a \sim 10^8\: {\rm
erg/cm}^3\: $  which lead to the inequality $\omega
\ll 10^{13}$. Hence in this case we are able to use standard
waveguides the width of which is of order 1~cm. Then we can
simplify the geometrical factor of the waveguide to the form
\begin{equation}
\label{lambda3} \lambda^2=\frac{H_0}{H_0-4\pi M_0}.
\end{equation}
This formula gives a remarkable possibility to change the angle of
the scissors
\begin{equation}
\label{angle} \theta=2\arctan\lambda
\end{equation}
by means of an external magnetic field applied along the
magnetization direction.

Moreover if to apply strong magnetic field $gH_0 \gg \omega$ or
$H_0 \gg 10^4 Oe $, then it follows from formula (\ref{lambda2}) that
\begin{equation}
\label{lambdaH} \lambda^2=\frac{H_0}{H_0+gK_a/M_0}.
\end{equation}
On this case also the effective angle of the structure can be
tuned by variation of the external magnetic field.

\section{Bound states and resonances}

First we review some general properties of the bound states which
can be derived by the standard methods of dealing with Laplace
operator with Dirichlet boundary conditions, the Dirichlet-Neumann
bracketing \cite{reed} and eigenvalue variation with respect to
the change of the domain \cite{kato}. The problem has two mirror
symmetries with respect to the axis of the angle $\theta$ which we
call scissor axis, and with respect to the axis of the larger
angle $180^o-\theta$ which we call the second axis. Using the
mentioned methods together with an eigenvalue-in-the-box estimate
analogous to that employed in \cite{avishai} we find that
 \begin{description}
 \item{(i)} every bound state is even w.r.t. the scissor axis,
 \vspace{-1.8ex}
 \item{(ii)} with respect to the second axis the bound states can
 have both parities which are alternating if the bound states are
 arranged according to their energies, \vspace{-1.8ex}
 \item{(iii)} as $\theta$ becomes smaller new bound states emerge
 from the continuum. The number $N$ of bound states satisfies the
 inequality $N\ge 2c\pi^{-1} (90^o/\theta)$ with $c= (1-2^{-2/3})
 ^{3/2}\approx 0.225$. While it is not good around $\theta=90^o$,
 where we know that $N=1$ from \cite{schult}, it is asymptotically
 exact as $\theta\to 0$,
 \vspace{-1.8ex}
 \item{(iv)} all the bound-state energies are monotonously
 increasing functions of $\theta$.
 \vspace{-1.8ex}
 \end{description}
The angle dependence of the bound-state energies has different
regimes. In the weak-coupling regime when the scissors are closing
and just passed the critical angle $\theta_c$ at which a new bound
state appeared, our numerical analysis shows that the binding
energy of the weakly-coupled states behaves as $\approx \pi^2-\gamma
(\theta_c-\theta)^2$ with some constant $\gamma$ which depends on the
particular state. On the other hand, strongly bound states
corresponding to a small $\theta$ are in the leading order
determined by the one dimensional potential well given by the
lowest transverse eigenvalue \cite{bento}. The second axis
determining the parity of the solution is then deep in the
classically forbidden region, so we can conclude that
 \begin{description}
 \item{(v)} as $\theta$ becomes smaller the bound states group into
 pairs with opposite parities and the energy gap between them is
 exponentially small as $\theta\to 0$.
 \end{description}
After these general results let us pass to the numerical
solution. We use three different methods. The most common among them
is the boundary integral method \cite{boundary}. In combination
with the above general results, it provides a rather complete
information about the discrete spectrum.

On the other hand, the boundary integral method tells us nothing
about the scattering problem in the scissor structure. We are
interested in particular in the scattering resonances associated
with quasibound states, which are characterized of complex
values of energy at which the analytically continued resolvent
has a pole singularity. A suitable method to treat this problem
is the exterior complex scaling. The method was suggested in the
seminal paper \cite{aguilar} and has developed into an efficient
computational tool -- see\cite{moiseev} and reference therein.
The use of exterior complex scaling for waveguide structures was
first proposed in \cite{duclos}, here we employ it in the form
presented in \cite{seba}. Before the proper scaling we pass to
right-angle scissors by means of the coordinate change
\begin{equation}
\biggl\{
\begin{array}{ccl}
x' & = & x \sin{\alpha} - y \cos{\alpha}\\ y' & = & y
\end{array}
\end{equation}
which takes the Hamiltonian to a unitarily equivalent operator
acting as
\begin{equation}
{\hat H} \Psi=\left(-\frac{\partial^2}{\partial
x'^2}-\frac{\partial^2}{\partial y'^2}
+2\cos{\alpha}\frac{\partial^2}{\partial x' \partial y'}\right)
\Psi
\end{equation}
Now we apply the scaling transformation to the longitudinal
variable in the structure arms which leaves the central area
unchanged, $x=g(X)$ and $y=g(Y)$, which yields the scaled
Hamiltonian
\begin{equation}
\label{Ham} {\hat H}=-\nabla \left[ \left(
\begin{array}{cc}
c_{11}(X,Y) & c_{12}(X,Y)\\ c_{21}(X,Y) & c_{22}(X,Y)\\
\end{array}
\right) \nabla \right] + U(X,Y)
\end{equation}
with

\begin{eqnarray*}
&& c_{11}(X,Y) = \frac{1}{g'^2(X)}\,, \qquad c_{12}(X,Y) =
-\frac{\cos{\alpha}}{g'(X) g'(Y)}\,, \\ && c_{21}(X,Y) =
-\frac{\cos{\alpha}}{g'(X) g'(Y)}\,, \qquad c_{12}(X,Y) =
\frac{1}{g'^2(Y)}\,,
\end{eqnarray*}
and
\begin{eqnarray*}
\lefteqn{ U(X,Y)=\frac{2 g'(X) g'''(X) - 5 g''^2(X)}{4 g'(X)^4}
} \\ && +\frac{2g'(Y) g'''(Y) - 5 g''^2(Y)}{4 g'(Y)^4}
+\frac{g''(X) g''(Y)}{4g'^2(X) g'^2(Y)}\, 2\cos{\alpha}
\end{eqnarray*}
The function $g(x)$ can be chosen, e.g., as
 $$ 
g(x) = \biggl\{
\begin{array}{lcl}
x &\quad\mathrm{if}\quad& |x| \le x_0 \\ \theta f(x)
&\quad\mathrm{if}\quad& |x|>x_0
\end{array}
 $$ 
with $x_0$ larger than the channel halfwidth and the
interpolating function $f(x)$ such that $f(x)=x$ for $|x|>2x_0$,
the function $g(x)$ is three times differentiable and the
inverse map $g^{-1}$ exists. As long as as the parameter
$\theta$ is real, the above transformation is a simple
coordinate change which does not change the spectrum. However,
if $\theta$ assumes complex values we observe a different
behavior in the discrete and continuous part typical in such
situations \cite{reed}: each branch of the continuous spectrum
of the operator (\ref{Ham}) is rotated into the complex plane
giving
 $$ 
\bigcup_{n=1}^\infty\{(n\pi)^2 + \theta^{-2}\langle 0,\infty)\}
 $$ 
for $d=1$. If $\Im{\rm m}\,\theta>0$ the rotated branches point
to the lower half plane and reveal parts of other sheets of the
Riemann surface of energy and we are able to see the resonance
poles as complex eigenvalues of the transformed operator; the
corresponding eigenfunctions are after the transformation
decaying at large distances, instead of the original growing
oscillations typical for Gamow functions.

Finally, the third method is based on application of
infitesimally small time-periodic perturbation. The bound states with
energies below the propagation subband ($E_b < E_0=\pi^2$) do not participate
in stationary transmission. However, it is possible to mix the bound state
$|b>$ with propagating state $|k>$ via a time-periodic perturbation
\begin{equation}
\label{timeperturb}
V(t)=V_0\cos(\omega t)
\end{equation}
provided that the matrix elements of the perturbation $<b|V|k> \neq 0$.
Such a possibility was demonstrated for the four-terminal's Hall junction
for electron transmission effected by a radiation field
\cite{bulgakov,bulgakov3}. Later the mixing of bound states with propagating
modes was also realized in microwave transmission \cite{bulgakov1}.
Here we similar to \cite{burmeister} use the time-periodic perturbation
(\ref{timeperturb}) as a probing
one to find the bound state energy by resonant features in the transmission
probability.

\section{Numerical results}

\begin{figure}[t]
\includegraphics[width=.4\textwidth]{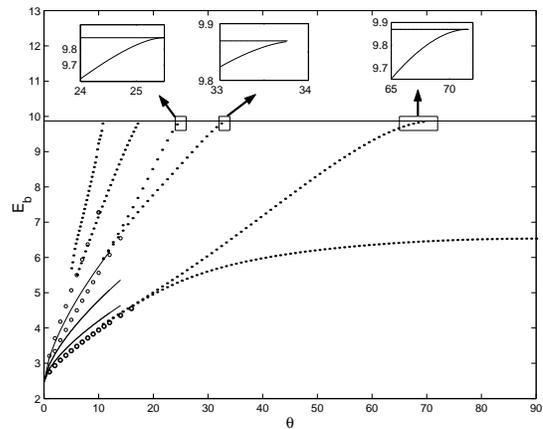}
\caption{Bound state energies for scissors structure
as a function of the interior angle
$\theta$. The complex scaling method data are shown by points. The
boundary integral method data are shown by circles. The asymptotic
formulas (\ref{limit}) with corresponding quantum numbers $m=1, 2,
3$  are given by thin solid lines. Insets above show blow up of asymptotic
behavior of the bound state energies in the vicinity of bottom
of propagation band $pi^2$.}
\label{scisfig2}
\end{figure}

Let us show the results of numerical analysis based on the methods
described above. First we plot the bound state energies as a
function of the scissor angle $\theta$.
The results of complex scale method are presented in Fig. \ref{scisfig2}
by points.

The results of boundary integral method
are shown in Fig. \ref{scisfig2} by circles. For the limit $\theta
\rightarrow 0$ the energies of bound states are derived in
\cite{carini1} and have the following form
Eq(22)
\begin{equation}\label{limit}
E_{nm}\rightarrow
\frac{\pi^2}{4}[n^2+(2n^2+m^2/4)\theta^{2/3}+...]
\end{equation}
where the quantum numbers $n, m = 1, 2, 3,...$. The factor $1/4$
in (\ref{limit}) takes into account that the length of an
inscribed rectangle has a length twicely more than in the case of
bent waveguide studied in \cite{carini,carini1}. Insets above the
figure show blow up of asymptotic behavior of the energies in the vicinity of
the edge energy band $E_0=\pi^2$. For all energies of bound states
the asymptotics are $\pi^2-\gamma (\theta_c-\theta)^2$ where $\gamma$
is a constant.

\begin{figure}[t]
\includegraphics[width=.4\textwidth]{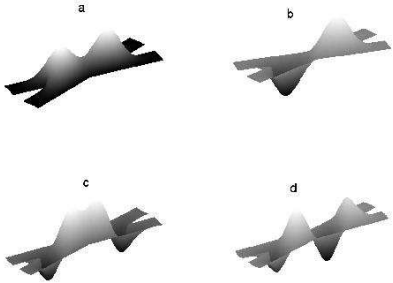}
\caption{The even-even and odd-even bound states for scissors
structure for $\theta = 30^o$.}
\label{scisfig3}
\end{figure}

As it was discussed in the Introduction
as the angle $\theta$ diminishes and the states become strongly
bound, the energy gap between the even and odd member of the pair
vanishes exponentially fast. In fact, one can see in Fig. \ref{scisfig2}
that the second bound state energy approaches to the first one, the fourth
bound state energy approaches to the third one, and so on.
In Fig. \ref{scisfig3} (a, b)
the first even-even and the second odd-even
bound states are shown. One can see that we have typical quantum
mechanical task of double well potential \cite{landau} in which
an energy distance between the first and second energy levels becomes
exponentially small with growth of a potential barrier between wells.
Fig. \ref{scisfig3} (c,d) demonstrates the next pair of the bound states
in the scissor structure.

\begin{figure}[t]
\includegraphics[width=.4\textwidth]{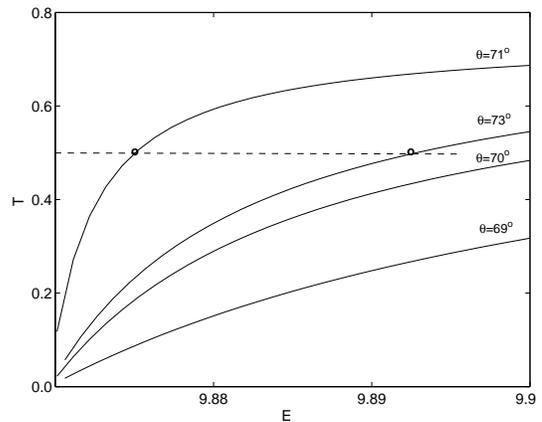}
\caption{The probability of transmission through the scissors structure
as function of $E=\lambda k^2$, eigen value of the Helmholz
equation (\ref{helmholz}). The transmission probability limits to zero
for $E\rightarrow E_0=\pi^2$. Circles show for which $E$ the transmission
probability equals the half.}
\label{scisfig4}
\end{figure}

\begin{figure}[t]
\includegraphics[width=.4\textwidth]{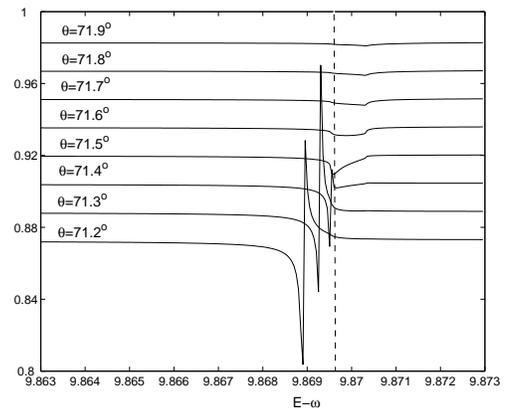}
\caption{Evolution of resonant features for the transmission through the
scissor structure caused by mixing the propagation state with the bound state
via the time-periodic perturbation (\ref{timeperturb}) as the angle of the
scissor $\theta$ increases. The evolution exactly follows to a blow up
shown in inset of Fig. \ref{scisfig2} for $\theta\rightarrow \theta_c - 0$.
For the angle exceeding $\theta_c$ resonant feature is decayed.
Dashed vertucal line shows the edge of the propagation band $E_0=\pi^2$.}
\label{scisfig5}
\end{figure}

    Usually the bound state transforms into
the quasibound state which displayes as resonant dip or peak for transmission
through the structure  as it was observed, for example, in \cite{exner5}.
However as one can see from Fig. \ref{scisfig4}
numerical computation of the transmission through the scissor's
structure does not show any resonant features for
$\theta > \theta_c\approx 71.5^o$. Also we used the time-periodic perturbation
method to search the quasi boun+ state above $\theta_c$. The results of
computation are in the vicinity of critical angle  $\theta_c$ are
shown in Fig. \ref{scisfig5}.

One can see that for $\theta < \theta_c$ we have resonant features
caused by mixing of propagating mode with the bound state of the
time-periodic perturbation. However for $\theta > \theta_c$ these
resonant features do not evolve but only decay with increasing of
the angle $\theta $. The small wiggle around the value 9.8704 is
an artefact of the computation which diminishes with the decrease
of $V_0$.

\begin{figure}[t]
\includegraphics[width=.4\textwidth]{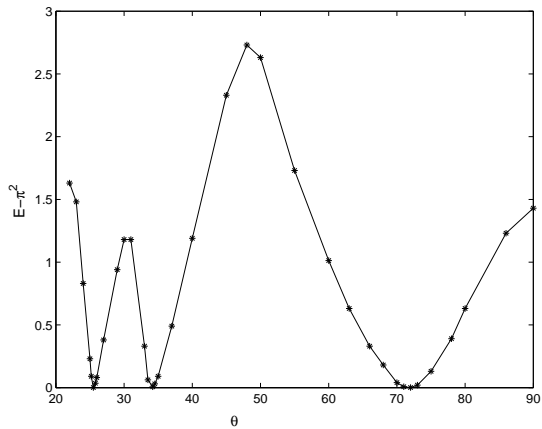}
\caption{Distances to the bottom of propagation band at which the transmission
takes the half (see Fig. \ref{scisfig4}) versus the angle of the scissors
waveguide.}
\label{scisfig6}
\end{figure}

Moreover for  $\theta$ approaching $\theta_c$
the transmission probability $T(E)$ undergoes the following feature
as shown in Fig. \ref{scisfig4}.
The less a value $|\theta -\theta_c|$ the more a slop of the transmission
versus $E$ in the vicinity of $E-\pi^2$. For  $\theta \rightarrow\theta_c$
the derivative $dT/dE\rightarrow \infty$.
If to plot $E-\pi^2$ at which
the transmission is reaching half as shown by circles in
Fig. \ref{scisfig4} versus the angle of scissor waveguide we obtain
remarkable curve shown in Fig. \ref{scisfig6}. We see that a value
$E-\pi^2$ equals zero just at the critical angles.

\acknowledgments This work has been partially by RFBR Grant
01-02-16077, GrantA1048101 of the GAS (P.Exner)
 and the Royal Swedish Academy of Sciences (A.Sadreev).

$^{*}$ e-mails: exner$@$ujf.cas.cz,
almsa$@$ifm.liu.se


\begin{thebibliography}{99}

\bibitem{ursell}
F. Ursell, Proc. Roy. Soc. {\bf 47}, 79 (1952)
\bibitem{roukes}
M.L. Roukes, A. Scherer, S.J. Allen, Jr., H.G. Graighead, R.M.
Ruthen, E.D.~Beebe, and J.P. Harbison, Phys. Rev. Lett. {\bf 59}, 3011 (1987)
\bibitem{timp}
G. Timp, H.U. Baranger, P. de Vegvar, J.E. Cunningham, R.E.
Howard, R.~Behringer, and M.M. Mankiewich, Phys. Rev. Lett. {\bf
60}, 2081 (1988)
\bibitem{exner1}
P. Exner and P. \v{S}eba, J. Math. Phys. {\bf 30}, 2574 (1989)
\bibitem{exner2}
P. Exner, P. \v{S}eba, and P. \v{S}\v{t}ov\'{\i}\v{c}ek, Czech. J.
Phys. {\bf B39}, 1181 (1989)
\bibitem{exner3}
P. Exner, Phys. Lett. {\bf A141}, 213 (1989)
\bibitem{schult}
R.L. Schult, D.G. Ravenhall, and H.W. Wyld, Phys. Rev. {\bf B39}
5476 (1989)
\bibitem{peeters}
F.M. Peeters, Superlatt. Microstruct. {\bf 6}, 217 (1989)
\bibitem{exner4}
P. Exner and P. \v{S}eba, Phys. Lett. {\bf A144}, 347 (1990)
\bibitem{avishai}
Y. Avishai, D. Bessis, B.G. Giraud, and G. Mantica, Phys. Rev.
{\bf B44}, 8028 (1992)
\bibitem{goldstone}
J. Goldstone and R.L. Jaffe, Phys. Rev. {\bf B45}, 14100 (1992)
\bibitem{carini}
J.P. Carini, J.T. Londergan, K. Mullen, and D.P. Murdock, Phys.
Rev. {\bf B46}, 15538 (1992)
\bibitem{carini1}
J.P. Carini, J.T. Londergan, K. Mullen, and D.P. Murdock, Phys.
Rev. {\bf B48}, 4503 (1993)
\bibitem{carini2}
J.P. Carini, J.T. Londergan, D.P. Murdock, D. Trinkle, C.S. Yung,
Phys. Rev. {\bf B55}, 9842 (1997)
\bibitem{bulgakov}
E.N. Bulgakov and A.F. Sadreev, JETP Letters, {\bf 66}, 431 (1997)
\bibitem{bulgakov1}
E.N. Bulgakov and A.F. Sadreev, Technical Phys. {\bf 46}, 1281
(2001)
\bibitem{londergan}
J.T.~Londergan, J.P.~Carini, D.P.~Murdock: {\em Binding and
Scattering in Two-Dimensional Systems. Applications to Quantum
Wires, Waveguides and Photonic Crystals}, Springer LNP~m60, Berlin
1999
\bibitem{bulgakov2}
E.N. Bulgakov and A.F. Sadreev, Europhys. Lett.
{57}, 198 (2002).
\bibitem{lax}
B. Lax, K.J. Button, {\em Microwave Ferrites and Ferrimagnetics}, N.Y.,
1962.
\bibitem{reed}
M.~Reed and B.~Simon: {\em Methods of Modern Mathematical Physics,
IV.~Analysis of Operators}, Academic Press, New York 1978;
Sec.~XIII.15.
\bibitem{kato}
T.~Kato: {\em Perturbation Theory for Linear Operators}, 2nd
edition, Springer, Berlin 1976; Sec.~VIII.6.
\bibitem{bento}
F.~Bentosela, P.~Exner, V.A.~Zagrebnov, Phys. Rev. B {\bf 57}, 1382
(1998)
\bibitem{boundary}
R.J.Riddel, Jr, J. Comp. Phys. {\bf 31}, 21, 42 (1979)
\bibitem{aguilar}
J.~Aguilar and J.--M.~Combes, Commun. Math. Phys.{\bf 22}, 269 (1971)
\bibitem{moiseev}
N.~Moiseyev,  Phys. Rep. {\bf 20}, 211 (1998)
\bibitem{duclos}
P.~Duclos, P.~Exner, P.~\v S\v tov\'\i\v cek, Ann. Inst.
H.~Poincar\'e: Phys. Th\'eor. {\bf 62}, 81 (1995)
\bibitem{seba}
P.~{\v S}eba, I.~Rotter, M.~Muller, E.~Persson, K.~Pichugin, Phys. Rev.
{\bf E61}, 66 (2000)
\bibitem{bulgakov3}
E.N. Bulgakov and A.F. Sadreev, JETP {\bf 87}, 1058 (1998)
\bibitem{burmeister}
G.Burmeister and K.Maschke, Phys. Rev. B{\bf 59}, 4612 (1999).

\bibitem{exner5}
P. Exner
and D. Krej$\check{c}$i$\check{r}$ik, J. Phys. A {\bf 32}, 4475 (1999)

\bibitem{landau}
L.D.Landau and E.M.Lifshitz, {\em Quantum Mechanics}, Pergamon Press,
N.Y., 1965
\end{thebibliography}
\end{document}